\documentclass[aps,prd,nofootinbib,showpacs]{revtex4-2}

\usepackage{amsmath,amssymb,bm,amsfonts}
\usepackage{color}

\newcommand{\be}{\begin{equation}}
\newcommand{\ee}{\end{equation}}
\newcommand{\bea}{\begin{eqnarray}}
\newcommand{\eea}{\end{eqnarray}}

\newcommand{\eq}[1]{Eq.~(\ref{eq:#1})}

\newcommand{\del}{\partial}

\newcommand{\bra}{\langle}
\newcommand{\ket}{\rangle}
\newcommand{\calO}{{\cal O}}

\newcommand{\eg}{{\it e.g.}}
\newcommand{\ie}{{\it i.e.}}

\newcommand{\bh}{black hole\ }

\newcommand{\SC}{superconductor}

\newcommand{\HSC}{holographic superconductor}
\newcommand{\HSCs}{holographic superconductors}

\bmdefine{\bmq}{{\bm{q}}}
\bmdefine{\bmk}{{\bm{k}}}
\bmdefine{\bmx}{{\bm{x}}}
\bmdefine{\bmy}{{\bm{y}}}
\bmdefine{\bmr}{{\bm{r}}}
\bmdefine{\bmnabla}{{\bm{\nabla}}}
\bmdefine{\bmA}{ \bm{A} }
\bmdefine{\bmD}{ \bm{D} }

\bmdefine{\bmPhi}{ \bm{\Phi} }
\bmdefine{\bmPsi}{ \bm{\Psi} }
\bmdefine{\bmcalO}{ \bm{\mathcal{O}} }
\bmdefine{\bmrho}{ \bm{\rho} }

\newcommand{\vecx}{\vec{x}}

\newcommand{\epsmu}{\epsilon_{\mu}}

\bmdefine{\bmg}{{\bm{g}}}
\bmdefine{\bmR}{{\bm{R}}}

\newcommand{\half}{\frac{1}{2}}

\newcommand{\la}{\lambda_\text{\tiny{GB}}}
\newcommand{\NGB}{N_\text{\tiny{GB}}}

\newcommand{\Atol}{A_t^{(0,\lambda)}}
\newcommand{\Atll}{A_t^{(2,\lambda)}}
\newcommand{\Psiol}{\Psi^{(1,\lambda)}}
\newcommand{\Psill}{\Psi^{(3,\lambda)}}

\newcommand{\mull}{\mu_{2\lambda}}

\newcommand{\eps}{\epsilon}

\newcommand{\Leff}{L_\text{eff}}

\begin{document}


\title{Comments on higher-derivative corrections in the AdS/CFT duality}
\author{Makoto Natsuume}
\email{makoto.natsuume@kek.jp}
\altaffiliation[Also at]{
Department of Particle and Nuclear Physics, 
SOKENDAI (The Graduate University for Advanced Studies), 1-1 Oho, 
Tsukuba, Ibaraki, 305-0801, Japan;
 Department of Mechanical Engineering, Mie University, 
Tsu, 514-8507, Japan.}
\affiliation{KEK Theory Center, Institute of Particle and Nuclear Studies, 
High Energy Accelerator Research Organization,
Tsukuba, Ibaraki, 305-0801, Japan}
%
\preprint{KEK-TH-2824} 
%
%
\begin{abstract}
We point out that some works on higher-derivative corrections in the AdS/CFT duality use inappropriate ``AdS/CFT dictionary." We illustrate the problem using a class of holographic superconductors in the Gauss-Bonnet black hole background. 
We also point out another problem in previous works. These modifications can change the qualitative behavior of physical quantities such as the spontaneous condensate. 
Whether the condensate increases or decreases under higher-derivative corrections depends on the system one considers, and there is no universality in this sense.

\vspace{18pt}
\small
\textit{Essay written for the Gravity Research Foundation 2026 Awards for Essays on Gravitation}
\end{abstract}
%

\maketitle



The AdS/CFT duality or the holographic duality claims that strongly-coupled field theories are equivalent to classical gravitational theories in AdS spacetime  \cite{Maldacena:1997re,Witten:1998qj,Witten:1998zw,Gubser:1998bc}. Strongly-coupled field theories are difficult to solve in general, so AdS/CFT is applied to solve strongly-coupled theories in ``real world." Examples are quark-gluon plasma, hadron physics, nonequilibrium physics, nonlinear physics, condensed-matter physics, quantum information, and so on (see, \eg, Refs.~\cite{CasalderreySolana:2011us,Natsuume:2014sfa,Ammon:2015wua,Zaanen:2015oix,Hartnoll:2016apf,Baggioli:2019rrs}). For that purpose, AdS/CFT has been used not only by AdS/CFT experts but also by people in different fields. Such collaborations among different fields produce many fruitful results.

The AdS/CFT duality consists of a set of rules called the ``AdS/CFT dictionary." We see one example below. But one should not take the dictionary at face value. These rules are not laws, rather they are empirical relations. 
In general, one needs to derive the AdS/CFT dictionary using a more fundamental relation such as the GKP-Witten relation. 
In particular, the standard dictionary assumes the asymptotically AdS geometry in the ``canonical" form:
\begin{align}
ds_5^2 \sim \left( \frac{r}{L} \right)^2 (-dt^2+d\vecx_3^2)+L^2 \frac{dr^2}{r^2}~, \quad (r\to\infty)~,
\label{eq:ads}
\end{align}
where $d\vecx_3^2=dx^2+dy^2+dz^2$.
If the asymptotic geometry does not take this form, one should be careful about the AdS/CFT dictionary. 

This is well-known among experts, but it is not often emphasized enough, 
and there are some confusions in the literature. The purpose of this note is to emphasize this point. 

As an example, we consider bulk 5-dimensional \HSCs\ in the Gauss-Bonnet (GB) AdS \bh background. The holographic superconductor \cite{Gubser:2008px,Hartnoll:2008vx,Hartnoll:2008kx} describes a \SC\ and is one of the most studied systems in AdS/CFT. The original \HSC\ uses Einstein gravity.
From the modern perspective, Einstein gravity is just the leading term in the effective theory expansion. Higher-derivative corrections such as GB gravity describe next order terms in the expansion.

The effect of higher-derivative corrections to \HSCs\ has been initiated in Ref.~\cite{Gregory:2009fj}. Since then, various works appeared. However, as we discuss below, previous works typically have 2 problems:
\begin{itemize}
\item
Previous works typically use the standard AdS/CFT dictionary. But the AdS/CFT dictionary gets higher-derivative corrections in the GB AdS \bh background. This will be clear if one rewrites the GB AdS \bh \eqref{eq:GB_bh} in the ``canonical form" \eqref{eq:ads}.
\item
Previous works implicitly assume that the kinetic term of the dual boundary theory has the canonical normalization, but this does not hold in general. In particular, the kinetic term normalization gets higher-derivative corrections.
\item
These modifications can change the qualitative behavior of physical quantities such as the spontaneous condensate under higher-derivative corrections.
\end{itemize}

The 5-dimensional Gauss-Bonnet AdS gravity is given by 
\begin{subequations}
\label{eq:GB}
\begin{align}
S = &\int d^{5}x \sqrt{-g}\left \{ R-2\Lambda+ \half\la L^2(R^2-4R_{AB}R^{AB}+R_{ABCD}R^{ABCD}) \right\}~,\\
\Lambda &=-\frac{6}{L^2}~.
%
\end{align}
\end{subequations}
Most literature on this system assumes $\la>0$. Following the tradition, we also assume $\la>0$. However, we study the $O(\la)$ corrections, so the qualitative behaviors of the GB \HSC\ depend on the sign of $\la$. 

The \bh solution \cite{Cai:2001dz} is given by
\begin{subequations}
\label{eq:GB_bh}
\begin{align}
%
ds_5^2 &= \left( \frac{r'}{L} \right)^2 \{ -f \NGB^2 dt^2+ d\vecx_3^2 \}+ L^2\frac{dr'^2}{r'^2f}~, \\
f &= \frac{1}{2\la} \left\{1-\sqrt{1-4\la \biggl\{1-\left(\frac{r_0'}{r'}\right)^4 \biggr\} } \right\}~.
%
\end{align}
\end{subequations}
The horizon is located at $r'=r_0'$. The constant $\NGB$ is used, \eg, in Ref.~\cite{Brigante:2007nu}.
In the limit $\la\to0$, the metric reduces to the 5-dimensional Schwarzschild-AdS (SAdS$_5$) black hole if $\NGB=1$.
The Hawking temperature is $\pi T=\NGB r_0'/L^2$.

\eq{GB_bh} is the exact solution of GB gravity \eqref{eq:GB}. However, we will keep $O(\la)$ terms only. Namely, we consider a small value of $\la$. This is because GB gravity describes just the first-order correction in the effective theory expansion. The action should have $O(R^3)$ and higher order terms in general. The effective theory expansion assumes that higher order terms are small compared with the leading Einstein gravity. Once those higher order terms become dominant, the effective theory expansion itself breaks down.

We consider the following \HSCs\ \cite{Herzog:2010vz} in the GB \bh background:
\begin{subequations}
\begin{align}
S_\text{m} &= -\frac{1}{g^2} \int d^5x \sqrt{-g} \biggl\{ \frac{1}{4}F_{MN}^2 + K|D_M\Psi|^2+V \biggr\}~,\\
K &= 1+A|\Psi|^2~,\quad
V= m^2 |\Psi|^2+B |\Psi|^4~.
%
\end{align}
\end{subequations}
Here, $A$ and $B$ are bulk parameters. The original \HSC\ is the $A=B=0$ case which we call the ``minimal" \HSC.
We consider the probe limit where the backreaction of matter fields onto the geometry is ignored. The background is the SAdS$_5$ \bh in Einstein gravity and the GB \bh in GB gravity.

At high temperature, the bulk matter equations admit a solution:
\begin{align}
A_t=\mu\biggl\{1-\left(\frac{r_0'}{r'}\right)^2 \biggr\}~,
A_i=0, \Psi=0.
%
\end{align}
But the $\Psi=0$ solution becomes unstable at the critical point and is replaced by a $\Psi\neq0$ solution. Then, the bulk field $\Psi$ plays the role of the order parameter. 

\subsection*{The asymptotic behavior of the GB AdS black hole}

Let us look at the asymptotic behavior of the GB AdS black hole  \eqref{eq:GB_bh}. 
The function $f$ of the GB \bh asymptotically behaves as
\begin{align}
f \sim  \frac{1}{2\la} \left\{1-\sqrt{1-4\la} \right\} \sim 1+\la~.
%
\end{align}
Then, the GB \bh behaves as
\begin{align}
ds_5^2 \sim \left( \frac{r'}{L} \right)^2 \{- (1+\la)\NGB^2dt^2+d\vecx_3^2\}+ \frac{L^2}{1+\la}\frac{dr'^2}{r'^2}~, 
\quad (r'\to\infty)~.
%
\end{align}
This  differs from the canonical form \eqref{eq:ads}, so one cannot directly apply the standard AdS/CFT dictionary below \eqref{eq:dict}. 
It is useful to rewrite the metric in the canonical form.

First, most previous works set $\NGB=1$. But, in this case, the boundary metric does not take the Minkowski metric $ds^2\propto -dt^2+d\vecx_3^2$. It is natural to choose $\NGB$ so that
\begin{align}
\NGB^2 \sim \frac{1}{1+\la} \sim 1-\la~.
%
\end{align}
Namely, our definition of the boundary time differs from previous analysis. 
Second, one often uses the convention 
\begin{align}
\Leff^2 \sim \frac{L^2}{1+\la}~.
%
\end{align}
Finally, make  the coordinate transformation $r'/L=r/\Leff$, or $r=\NGB r'$. Then, the asymptotic behavior becomes
\begin{align}
ds_5^2 \sim \left( \frac{r}{\Leff} \right)^2 (- dt^2+d\vecx_3^2)+\Leff^2 \frac{dr^2}{r^2}~, 
\quad (r\to\infty)~.
%
\end{align}
The metric now takes the canonical form \eqref{eq:ads}, but $L$ is replaced by $\Leff$: this becomes important below.

\subsection*{Problems of previous works}

Previous works typically have the following 2 problems:
\begin{enumerate}
\item
Problem of the AdS/CFT dictionary:
In the canonical form \eqref{eq:ads}, the asymptotic behavior of the bulk scalar field is given by
\begin{subequations}
\begin{align}
\Psi &\sim \psi_- \left( \frac{L}{r}\right)^{\Delta_-} + \psi_+ \left( \frac{L}{r}\right)^{\Delta_+}~,
\quad (r\to\infty)~, \\
\Delta_\pm &= 2\pm\sqrt{4+m^2L^2}~.
%
\end{align}
\end{subequations}
Then, the standard procedure like the GKP-Witten relation gives
\begin{subequations}
\label{eq:dict}
\begin{align}
\bra\calO\ket &= \frac{\Delta_+-\Delta_-}{g^2L}\psi_+~, \\
J &=\psi_-~,
%
\end{align}
\end{subequations}
where $\bra\calO\ket$ is the order parameter (condensate) and $J$ is its source. This is one example of the ``AdS/CFT dictionary."

An important point is that the normalization of $\bra\calO\ket$ is automatically fixed, and one cannot choose it. This point is often ignored in previous literature but becomes important for higher-derivative corrections as we see below. 

For the GB black hole, $L$ is replaced by $\Leff$, so the dictionary is modified as
\begin{subequations}
\label{eq:correct}
\begin{align}
\Psi &\sim \psi_- \left( \frac{\Leff}{r}\right)^{\Delta_-} + \psi_+ \left( \frac{\Leff}{r}\right)^{\Delta_+}~,
\quad (r\to\infty)~, \\
\Delta_\pm &= 2\pm\sqrt{4+m^2\Leff^2}~,\\
\bra\calO\ket &= \frac{\Delta_+-\Delta_-}{g^2\Leff}\psi_+~.
%
\end{align}
\end{subequations}
%
%
The AdS/CFT dictionary gets $O(\la)$ corrections because $\Leff^2=\NGB^2L^2\sim (1-\la)L^2$.

$\Delta_\pm$ are called the scaling dimensions. The scaling dimensions are also $\la$-dependent.
%
%
If one fixes the scaling dimensions, the overall factor $(\Delta_+-\Delta_-)$ remains the same. But if one fixes the bulk scalar mass $m$ \cite{Gregory:2009fj}, one gets an $O(\la)$ correction from the overall factor as well.

We illustrate the modification of the AdS/CFT dictionary for the bulk scalar field, but one needs similar modifications for  the charge density $\bra J^t\ket$ and the current density $\bra J^i\ket$. 


Here, we use the coordinate transformation to obtain the AdS/CFT dictionary, but there are several ways. A simpler way is to notice that the ``pure" AdS geometry \eqref{eq:ads} remains the solution of GB gravity but $L$ is replaced by $\Leff$ due to higher-derivative corrections. Then, it is clear that one needs to replace $L$ by $\Leff$. 
Also, it is always safe to start with a fundamental relation such as the GKP-Witten relation.
Namely, one can derive the same dictionary from the GKP-Witten relation in the GB AdS black hole background \eqref{eq:GB_bh}.

\item
Problem of the canonical normalization of the dual Ginzburg-Landau theory:
From the dual point of view, the dual theory should be described by the Ginzburg-Landau (GL) free energy:
\begin{align}
f &= c |\del_i\psi|^2-a|\psi|^2+\frac{b}{2}|\psi|^4~,
%
\end{align}
where $\psi=\bra\calO\ket$. In the high-temperature phase, there is no condensate $\psi=0$, but in the low-temperature phase $|\psi|\neq0$.
Previous works compute the condensate from the potential terms and did not consider the kinetic term. Namely, they compute
\begin{align}
|\psi|^2 = \frac{a}{b}~.
%
\end{align}
A hidden assumption in previous works is that the kinetic term of the dual GL theory has the canonical normalization $c=1$, but there is no guarantee. In fact, the kinetic term typically does not have the canonical normalization, \ie, $c\neq1$ \cite{first}. In such a case, one should rescale 
$|\tilde{\psi}|^2=c|\psi|^2$,
and the real condensate is 
\begin{align}
|\tilde{\psi}|^2= \frac{a}{b}c~.
\label{eq:real_condensate}
\end{align}
The normalization $c$ gets an $O(\la)$ correction, so one should take it account.
Namely, the condensate cannot be determined simply by solving a homogeneous problem.
\end{enumerate}
As we saw, there are at least 3 additional places where $\la$ appears:
\begin{enumerate}
\item
The factor $\NGB$ in the metric,
\item
The factors $\Leff$ in the dictionary \eqref{eq:correct},
\item
 The GL kinetic term normalization $c$ \eqref{eq:real_condensate},
\end{enumerate}
but they are mostly ignored in previous literature. 

Usually, these modifications 
 just change overall factors and are relatively harmless. But these factors get higher-derivative corrections, so one needs to take them into account. The modifications can change the qualitative behavior of physical quantities such as the spontaneous condensate.

For example, we previously obtained the analytic solution for a bulk 5-dimensional minimal \HSC\ in the GB \bh background \cite{second}. 
If one ignores these factors,  the condensate $|\psi|^2$ ``decreases"  for the system (in the sense below). 
However, if one takes these factors into account,  these factors add up together and the condensate $|\tilde{\psi}|^2$ ``increases."
Namely, these factors overwhelm the other background geometry effects.

However, there is no reason to believe that this is always the case. 
As a further example, consider ``nonminimal" \HSCs\ with $A,B\neq0$. As in our previous analysis \cite{second}, we consider the case $\Delta_\pm=2$ because we can construct the analytic solution (see,\eg, Refs.~\cite{Herzog:2009ci,Herzog:2010vz,Natsuume:2018yrg} for the other analytic solutions of \HSCs).

 A \HSC\ has 2 dimensionful control parameters, temperature $T$ and chemical potential $\mu$, so the system is parametrized by a dimensionless parameter $\mu/T$. We fix $T$ and vary $\mu$. 
 Then, the deviation from the critical point is expressed by $\epsmu=\mu-\mu_c$, where $\mu_c$ is the critical point. For simplicity, we also set the AdS radius $L=1$ and we work in the unit $\pi T=1$.

The details of the computation can be found in the arXiv version of Ref.~\cite{second}. (App.~B which is added after publication) The critical point is given by
\begin{align}
\left(\frac{\mu}{\pi T} \right)_c \approx 2+1.682\la~.
%
\end{align}
The critical chemical potential $\mu_c$ becomes higher, or the critical temperature $T_c$ becomes lower due to the higher-derivative corrections. 
The parameters $A,B$ do not change the critical point.

Again, if one does not take the above 3 factors into account, the condensate $|\psi|^2$ ``decreases" when $A,B>0$.
But the canonically normalized condensate $|\tilde{\psi}|^2$ is given by
\begin{align}
|\tilde{\psi}|^2 \approx \frac{6}{1+4A+4B}\epsmu+ \la\frac{1.826-22.88A+2.306 B}{(1+4A+4B)^2}\epsmu~. 
%
\end{align}
For the minimal \HSC\ with $A=B=0$, the condensate increases since the $O(\la)$ correction is positive as mentioned above. Also, the condensate increases when $B>0$. But the condensate can decrease by choosing $A$ appropriately. 
The condensate can increase or decrease depending on the system, and there is no universality.

As we argued, most previous works do not consider the issues raised in this note, so their results must be reexamined. We illustrate the issues using \HSCs\ as examples, but the other works on higher-derivative corrections, in particular works using the GB AdS black hole,  may have the same problems. We hope that this note is helpful for future works on higher-derivative corrections in the AdS/CFT duality.

\begin{acknowledgments}

I would like to thank Takashi Okamura for his continuous suggestions and interest throughout the work.
This research was supported in part by a Grant-in-Aid for Scientific Research (25K07291) from the Ministry of Education, Culture, Sports, Science and Technology, Japan. 

\end{acknowledgments}

\end{document}